\documentstyle[11pt,preprint,amssymb,amsbsy,aps,epsf]{revtex}

\begin {document}
\title{Scattering of a particle with spin by atomic chain
as null test
of T-violating P-even magnetism}
\author{ S.L. Cherkas}
\address{ Institute of Nuclear Problems
220050 Minsk, Belarus}
\maketitle
\begin{abstract}
T-odd P-even long-range
electromagnetic
interaction of a particle
of spin 1/2 with the nucleus is considered.
Though matrix element of the interaction is zero for the
particles on mass shell, nevertheless, null
test exists for the interaction. The test consists in
measuring of the spin-dependent  T-odd P-even
forward elastic scattering amplitude of a particle of
spin 1/2 by atomic chain (axis) in a crystall.
\end{abstract}
\bigskip

In connection with the direct observation of
time-reversal symmetry
violation in
the system  of $K^0-\bar K^0 $ mesons
\cite {cplear}
it would be interesting to detect
T-non-invariance in other nuclear or atomic systems.
It is necessary to distinguish P- T- odd interaction from
P-even T-odd one. While on the
strength constants of the first type interactions
there are rather rigid restrictions, obtained from  dipole
moment of atoms and particles measuring, restrictions on
the constants of P-even T-odd interactions
are not so strong.
As it is known, zero test of the last kind interactions
is observation  of
$ \sim (\boldsymbol \sigma\times \boldsymbol n
\cdot \boldsymbol S) (\boldsymbol n
\cdot \boldsymbol S) $  term
in the forward elastic scattering amplitude of
a particle of spin $1/2 $
by a particle with a spin $S\ge 1 $ \cite{bar1,conzet,beyer,cher},
where $ \boldsymbol n $ is a unit vector in a direction
of an incident particle momentum.
Relevant
experiments
were carried out \cite {exp} and
plan to be performed on a superconducting synchrotron COSY
\cite {COSY}.

In this work new null test of T-invariance is
presented, which
consists in measuring of the T-odd forward
elastic scattering
amplitude of a
particle with spin by an atomic chain (axis) or a plane.

The motion of a particle with spin in matter can be
described with the help of
refraction
index, depending on a spin \cite {barbook}:
$
{ \mathcal N} =1 +
{ \displaystyle \frac {2\pi\wp} {p^2}} f (0), ~~~
$
where
$f (0) $ is forward scattering amplitude of a particle
with spin by  a
scattering
centre of a medium, $p $ is a wave number
of a particle,
$ \wp $ is the density of scatterers in the matter.
The dependence of refraction
index on the orientation of spin leads to the  experimentally
observable effects: absorption of a particles beam,
depending on the orientation
of particle spin (dichroism),
rotation
and oscillation of a particle spin \cite {bar1983,kost}.

In passage of a particle under small angles to the
atomic chains or planes the scattering occurs not on the
separate atom, but on the whole atomic chain or plane.

For the angles,
less than Lindhard angle,
negatively charged particle
is captured by an axis or
a plane and goes in its vicinity, i.e.
the channelling phenomenon
arises \cite {Lin} (positively charged particle is
localised between neighbour atomic
planes).
Channelling and
radiation under  channelling
were actively studied earlier \cite {Lin,barbook2,lug}.

But we are interesting the case, when a particle
goes under angle more  than
Lindhard one, at which the channelling, begins.
However
the angle is
so small
( less than $\frac{R_a}{d}$, $R_a$ is atomic radius, $d$ is
interatomic distance in chain),
that the particle motion
can be
described by  the averaged axis or plane potential.

In this case particle refraction index
is also determined by the formula above,
but the amplitude should  be read as denoting particle
forward elastic scattering
amplitude by the whole
axis or plane (accounted for unit axis length or unit area of the plane).

\begin{figure}[h]
\epsfxsize=4.0 cm
\hspace{2 cm}
\vspace{1 cm}
\epsfbox[0 300 400 450]{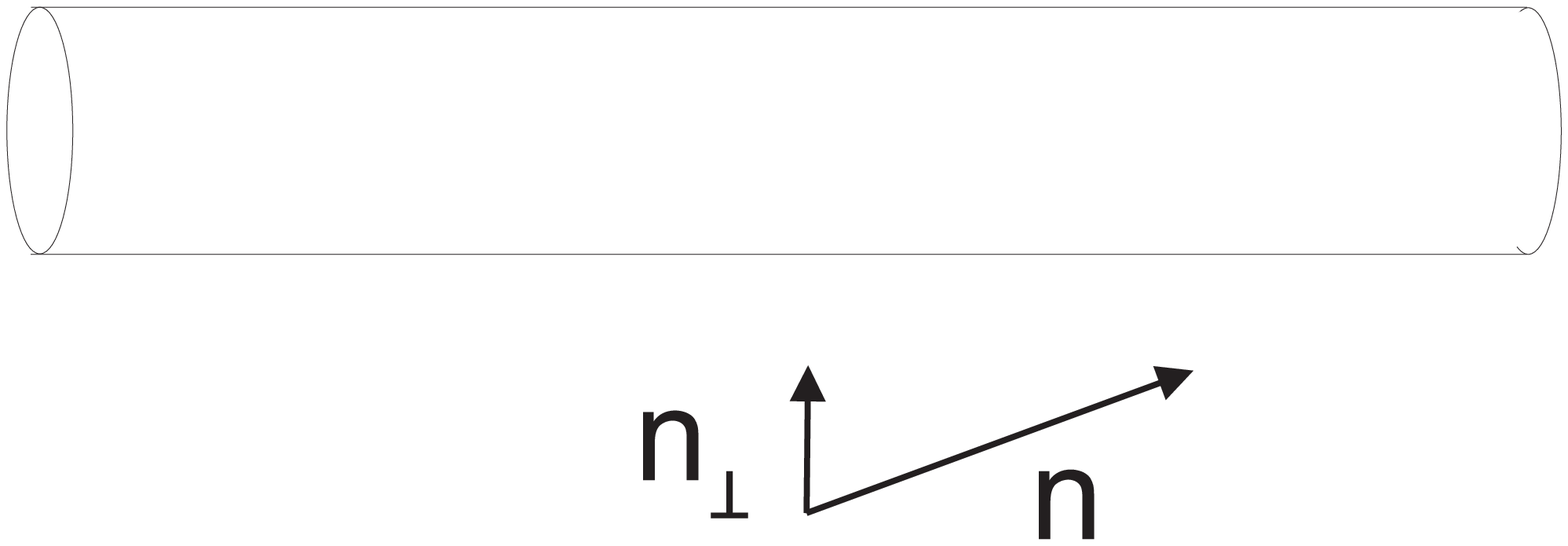}
\caption{Particle scattering by the atomic axis.}
\label{Fig1}
\end{figure}

Accordingly, a density of scatterers
should be understand as
two-dimensional density of cross points of axes,
with the plane being perpendicular to the axes, or one-dimensional
density of
cross points of planes by their normal line.
Let's designate the
direction of an axes or direction of planes
normal line
by the unit vector $ \boldsymbol a $
(fig.\ref{Fig1})
and
consider
spin
structure of the scattering amplitude at zero
angle of a particle with spin
$\boldsymbol S$
by an axis or a
plane:
\begin {eqnarray}
F (0) =A+B (\boldsymbol S \boldsymbol n) +B_1
( \boldsymbol S\cdot\boldsymbol a)
( \boldsymbol n \cdot \boldsymbol a) +
C_1 (\boldsymbol S\times \boldsymbol a \cdot \boldsymbol n)
( \boldsymbol a\cdot \boldsymbol n) +
A_1 (\boldsymbol S \cdot \boldsymbol n) ^2 +
\nonumber
\\
+A_2 (\boldsymbol S \cdot \boldsymbol a) ^2
+
A_3 (\boldsymbol S\cdot \boldsymbol a)
( \boldsymbol S\cdot \boldsymbol n) (\boldsymbol n\cdot \boldsymbol a) +
C_2 (\boldsymbol S\times \boldsymbol a \cdot \boldsymbol n)
( \boldsymbol S\cdot \boldsymbol a) + \dots,
\label {ampl}
\end {eqnarray}
$ \boldsymbol n $ being the unit vector in a direction
 of incident particle momentum.
For a particle with a spin 1/2 only three first terms should be
in the expression (\ref{ampl}), the remained terms occur
for particles with spins greater than 1/2.
The coefficients $A_n, B_n\cdots$ are even functions of
$ (\boldsymbol n\cdot \boldsymbol a) $.
From the view-point of fundamental
symmetries chesking T-violating/P-conserving $ \sim (\boldsymbol S\times
\boldsymbol a \cdot \boldsymbol n)
( \boldsymbol a\cdot \boldsymbol n) $ and
T-/P-violating $ \sim (\boldsymbol S\times \boldsymbol a
\cdot \boldsymbol n)
( \boldsymbol S\cdot \boldsymbol a) $ terms are interesting.
Experimental observation of these terms gives novel
null test of T-invariance.
It turns out to be,
that the first term can arise from
the T-odd/P-even particle interaction
\[
V_T(\boldsymbol r)=(\boldsymbol S\times
\hat {\boldsymbol p} \cdot \boldsymbol r)
u^T(r)
(\boldsymbol r \cdot \hat {\boldsymbol p}) +
( \hat {\boldsymbol p}\cdot\boldsymbol r)u^T(r)
(\boldsymbol r\cdot \boldsymbol S\times
\hat {\boldsymbol p})~,
\]
and the second term from the T-/P-odd
\[
V_{PT}(\boldsymbol r)=(
\boldsymbol S\times  \hat {\boldsymbol p}
\cdot
\boldsymbol r)u^{PT}(r)
(\boldsymbol r \cdot \boldsymbol S) +
(\boldsymbol r\cdot \boldsymbol S)u^{PT}(r)
(\boldsymbol r\cdot \boldsymbol S\times
\hat {\boldsymbol p})~,
\]
one.
$\hat {\boldsymbol p}$  designates operator of particle
momentum and
$u^{PT}(r)$, $u^{T}(r)$ being some radial functions.
These interactions do not lead to the observable effects in
ordinary elastic scattering: evaluation of
T-odd
amplitude in the Born approximation
gives zero. The same result will be if one
calculate the
scattering amplitude by chain of atoms in the Born approximation.
But taking into account the perturbation
of the plane waves
by the cylindrical
Coulomb potential of an axis in the distorted wave approximation
\cite {Dav} leads to the
non zero
T-odd amplitude.

\begin {figure} [h]
\epsfxsize=7.0 cm
\vspace {1 cm}
\hspace {4.5 cm}
\epsfbox [0 50 400 200] {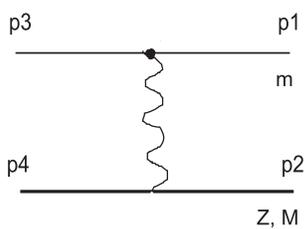}
\vspace {0.5 cm}
\caption{The diagram of one photon exchange.
The T-violating P-even vertex is designated by a black circle.
}
\label {Fig2}
\end {figure}

Let us consider T-violating interaction arising from one
photon exchange.
T-non-invariant  parity conserving electromagnetic
interaction
vertex
of a particle of spin $1/2$ with the electromagnetic field can be written
as \cite {t1,haff}
\begin{equation}
\Gamma^\mu_T=\frac{e\,{\mathfrak e}_T}{m^2}
\big((Pq)\gamma^\mu-(\gamma q)P^\mu\big)-
\frac{e\,{\mathfrak m}_T}{2m^3}(Pq)\sigma^{\mu\nu}q_\nu~,
\label{vert}
\end{equation}
where $P=p_3+p_1$, $q=p_3-p_1$  (fig.\ref{Fig2}).
Considering the spinless charged nuclei we take
ordinary electromagnetic
vertex in the form $Ze P_\mu $ \cite {ber}, $k=p_4+p_2 $ .
After the application of ordinary diagramm
technique, we obtain appropriate matrix  element
due to one photon exchange:
\begin{equation}
{\mathcal M}=\frac{Ze^2\mbox{i}}{m^2}\biggl\{
{\mathfrak e}_T\big((Pq)\gamma^\mu_{31}-(\gamma_{31} q)P^\mu\big)-
\frac{e\,{\mathfrak m}_T}{2m}(Pq)\sigma^{\mu\nu}_{31}q_\nu
\biggr\}D_{\mu\eta}(q)k^\eta~~,
\label{m}
\end{equation}
where $ \gamma ^\mu _ {31} \equiv\bar u_3 (p_3)
\gamma ^\mu u_1 (p_1) $, \dots
  and
$D _ {\mu\nu} (q) =4\pi g _ {\mu\nu} /q^2 $ is a photon propagator.
Setting $p_1=p $, $p_3=p+q $, $q = (0, \boldsymbol q) $
and substituting
$
u (p) =
\left (
\begin {array}{c}
\sqrt {\varepsilon+m} \; \phi \\
(\varepsilon+m)^{-1/2} \; (\boldsymbol \sigma \boldsymbol p) \; \phi
\end {array}
\right)~,
$
($ \phi $ is spin wave
function of a particle) into (\ref{m}) we find the T-odd
scattering amplitude of a particle by resting
nucleus
for the small momentum
transferred
$ \boldsymbol q $:
\begin{equation}
f(\boldsymbol q)=-\frac{\mathcal M}{8\pi M}=-\frac{2Ze^2}{m^2}
\left(\frac{{\mathfrak m}_T}{m}+\frac{{\mathfrak e}_T}{\varepsilon +m}
\right)
\frac{(\boldsymbol p \boldsymbol q)(
\boldsymbol \sigma\times \boldsymbol p\cdot
\boldsymbol q)}{q^2}.
\label{f}
\end{equation}
In evaluation of the amplitude we consider
the particle being
on mass shell everywhere
except for the
term $(\boldsymbol p\cdot \boldsymbol q)$.
If a particle is completely on
mass shell,
$(\boldsymbol p\cdot \boldsymbol q) =0$ and the amplitude is
zero.
Dependence of the amplitude on
momentum transferred
 $ \boldsymbol q $ is the same as
for elastic scattering amplitude of
magnetic dipoles. So, it turns out to be that the
interaction is
long-range one.

Improper conclusion  have been made
earlier \cite {haff}
( repeated then in the monograph \cite {blin})
about non-existence of  long-range (i.e.
decreasing as
$ {\displaystyle \frac {1} {r^3}} $
with distance
or weaker \cite {Dau})
T-odd potential.

We can consider scattering of a particle in
a
framework of the
Schrodinger equation with the relativistic mass \cite{ber}:
\begin {equation}
\label {sr}
( \nabla^2 + p^2) \Psi ({\boldsymbol r})
=2\varepsilon V (\boldsymbol r) \Psi (\boldsymbol r).
\end {equation}

In the Born approximation
the amplitude (\ref {f}) can be obtained
from the T-odd interaction, depending on the energy:
\begin {equation}
{V} _T (\boldsymbol r) = \frac {3Ze^2} {2 \varepsilon m^2}
\left (\frac {{\mathfrak m} _T} {m} + \frac {{\mathfrak e} _T} {\varepsilon +m}
\right) \bigl ((
\hat {\boldsymbol p} \boldsymbol r) \frac {1} {r^5} (
\boldsymbol r\cdot\boldsymbol\sigma\times \hat {\boldsymbol p}) + (
\boldsymbol \sigma\times
\hat {\boldsymbol p} \cdot \boldsymbol r)
\frac {1} {r^5} (\boldsymbol r
\hat {\boldsymbol p}) \bigr) ~,
\end {equation}
wich can be presented in the form:
\[
{V}_T(\boldsymbol r)=-\frac{e}{2 \varepsilon m^2}
\left(\frac{{\mathfrak m}_T}{m}+\frac{{\mathfrak e}_T}{\varepsilon +m}
\right)
\bigl(
\hat {\boldsymbol
p}\cdot\{\boldsymbol \nabla\otimes \boldsymbol E(
\boldsymbol r)\}\cdot (\boldsymbol \sigma\times
\hat {\boldsymbol p})+
(\boldsymbol \sigma\times
\hat {\boldsymbol p})\cdot\{
\boldsymbol \nabla\otimes \boldsymbol E(
\boldsymbol r)\}\cdot
\hat {\boldsymbol p}
\bigr)~,
\]
where $ \boldsymbol E (\boldsymbol r) = -
\boldsymbol \nabla \phi (\boldsymbol r) =Ze
{ \displaystyle \frac {
\boldsymbol r} {r^3}} $ is a strength of the electric
field created by the nucleus at the point $ \boldsymbol r $. $ \otimes $
means a direct vector product.
Till now we did not take into consideration
screening of the nucleus field by the
electrons. To remedy this
we take the nucleus  electric
potential in the form
${\displaystyle \phi(r)=\frac{Ze}{r}\mbox{Erfc}
\left(\sqrt{\frac{3}{2}}\frac{r}{R_a}\right)}$,
where
$\mbox{Erfc}(x)=\frac{2}{\sqrt \pi}\int\limits_{x}^{\infty}
\mbox{e}^{-t^2}d t$ and $R_a $ is atomic radius.
Now, as it is usually done in the theory of channelling,
we find
the T-odd average potential
of the atomic chain, with the thermal vibrations  of atoms
taken into account:
\begin{eqnarray}
v_T(
\boldsymbol
\rho)=\frac{1}{d}\int_{-\infty}^{+\infty}\int
{V_T}(
\boldsymbol r-\boldsymbol R)\left(\frac{3}{2\pi}\right)^{3/2}
R_T^{-3}\exp\left(-\frac{3R^2}{2R_T^2}\right)d^3
\boldsymbol Rdz~,
\nonumber
\\
v_T(\boldsymbol \rho)=\frac{e}{2m^2 \varepsilon  }
\left(\frac{{\mathfrak m}_T}{m}+\frac{{\mathfrak e}_T}
{\varepsilon +m}
\right)
\bigl(
(
\boldsymbol \sigma \times
\hat {\boldsymbol p} \cdot
\boldsymbol \rho)2g^\prime(\rho^2)(\boldsymbol \rho
\hat {\boldsymbol p})+
(\hat {\boldsymbol p}
\boldsymbol \rho)2g^\prime(\rho^2)(
\boldsymbol \rho\cdot
\boldsymbol \sigma \times
\hat {\boldsymbol p})
\label{pot2}
\\
\nonumber
-(\boldsymbol \sigma\times
\hat {\boldsymbol p} \cdot \boldsymbol a)
g(\rho^2)(\boldsymbol a
\hat {\boldsymbol p})-
(\boldsymbol a
\hat {\boldsymbol p})
g(\rho^2)(\boldsymbol \sigma\times
\hat {\boldsymbol p} \cdot
\boldsymbol a)
\bigr)~,
\end{eqnarray}
where $R_T^2 $ is
root-mean-square amplitude of an atom thermal vibrations.
Function
$
g(\rho^2)=-\frac{1}{\rho}\frac{d \Phi(\rho)}{d \rho}
$
is expressed through the ordinary
T-even axis potential
\begin{equation}
\Phi(\rho)=\frac{Ze}{d}\left(E_1\left(\frac{3\rho^2}{2R_T^2+2R_a^2}\right)-
E_1\left(\frac{3\rho^2}{2R_T^2}\right)
\right)~,
\end{equation}
obtained by the averaging of the screened  Coulomb
 potential
$\phi(r)$ in the manner above.
$
E_1(x)=\int\limits_{1}^{\infty}t^{-1}e^{-xt} dt
$ is exponential integral.

Thus,
the particle
interaction with the axis consists of
T-odd interaction
$v_T (\rho) $
and T-even one
$v(\rho)=e\Phi(\rho)$.

T-odd amplitude per unit length of an axis
can be calculated in the distored waves approximation \cite {Dav}:
\begin{equation}
F_T(0)=-\frac{\varepsilon}{2\pi} \frac{1}{d}
\int\limits_{-d/2}^{d/2}dz\int
d^2\boldsymbol \rho\Psi^{*(-)}(\boldsymbol r)v_T(
\boldsymbol \rho)\Psi^{(+)}(\boldsymbol r).
\label{dis}
\end{equation}
Plane wave distortion
by the Coulomb axis potential
can be easy taken into account
in terms of the
 eikonal approximation \cite {Dau}
using Schrodinger
equation
(\ref {sr}):
\begin{equation}
\Psi^{(+)}(\boldsymbol r)=exp\left(\mbox{i}
\boldsymbol p\boldsymbol r-\mbox{i}\frac{\varepsilon}{p}
\delta(\rho^2,(\boldsymbol n_\bot
\boldsymbol \rho))
\right),~~
\Psi^{(-)}(\boldsymbol r)=exp\left(\mbox{i}
\boldsymbol p\boldsymbol r+\mbox{i}\frac{\varepsilon}{p}
\delta(\rho^2,-(\boldsymbol n_\bot\boldsymbol \rho))
\right),
\label{voln}
\end{equation}
where
$
\delta(\rho^2,(\boldsymbol n_\bot
\boldsymbol \rho))=\int\limits_{-\infty}^{(\boldsymbol n_\bot
\boldsymbol \rho)}
v(\mid \boldsymbol \rho-
\boldsymbol n_\bot(\boldsymbol n_\bot\boldsymbol \rho)+
\boldsymbol n_\bot t\mid)d t,
$
$ \boldsymbol n_\bot $ is perpendicular to the axis component
of a
unit vector in the particle momentum direction.
Note, that
$ \delta $ depends from $ \boldsymbol \rho $ only through combinations
$ \rho^2 $ and $ (\boldsymbol n_\bot \boldsymbol \rho) $,
that is corollary of the cylindrical
symmetry of the system considered.
Apparently, the same T-odd effect is
to exist for neutrons also. In this case
  plane wave distorting by
cylindrical potential originates from  the
neutron magnetic moment
interaction
with the charged nuclei or from the
strong neutron interaction with nuclei.
Substituting wave functions (\ref {voln}) and
T-odd potential (\ref {pot2})
into
the equation (\ref {dis})
we obtain:
\begin{eqnarray}
F_T(0)=\frac{Ze^2\varepsilon}{2\pi m^2 d}
\left(\frac{{\mathfrak m}_T}{m}+\frac{{\mathfrak e}_T}{
\varepsilon +m}\right)
(
\boldsymbol n \boldsymbol a)(
\boldsymbol \sigma\cdot \boldsymbol a \times \boldsymbol n)
\nonumber
\\
\times \frac{1}{n_\bot^2}
\int e^{-\mbox{\scriptsize i}\frac{\varepsilon}{p}(\delta(X,Y)+\delta(X,-Y))}
\biggl\{\left(2YX\frac{\partial \delta(X,Y)}{\partial X}
+Y^2\frac{\partial \delta(X,Y)}{\partial Y}\right)2g^\prime(X)
\nonumber \\ +
\left(2Y\frac{\partial \delta(X,Y)}{\partial X}+n_\bot^2
\frac{\partial \delta(X,Y)}{\partial Y}\right)g(X)
\biggr\}d^2\boldsymbol\rho~~,
\label{end}
\end{eqnarray}
where $X =\rho^2 $ and $Y = (\boldsymbol \rho \cdot
\boldsymbol n_\bot) $.

\begin {figure} [h]
\epsfxsize=7.0 cm
\epsfbox {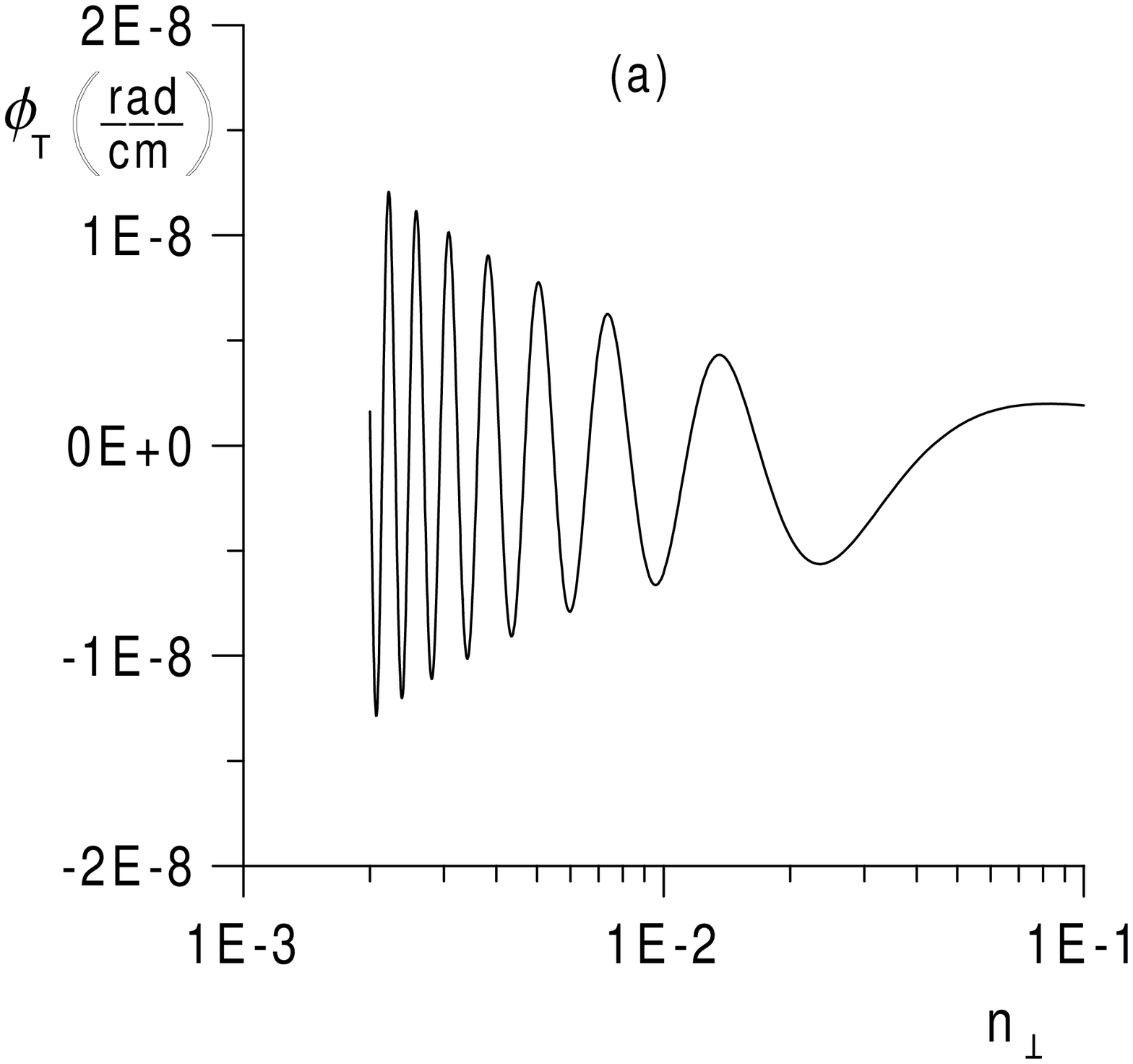}
\epsfxsize=7.0 cm
\epsfbox {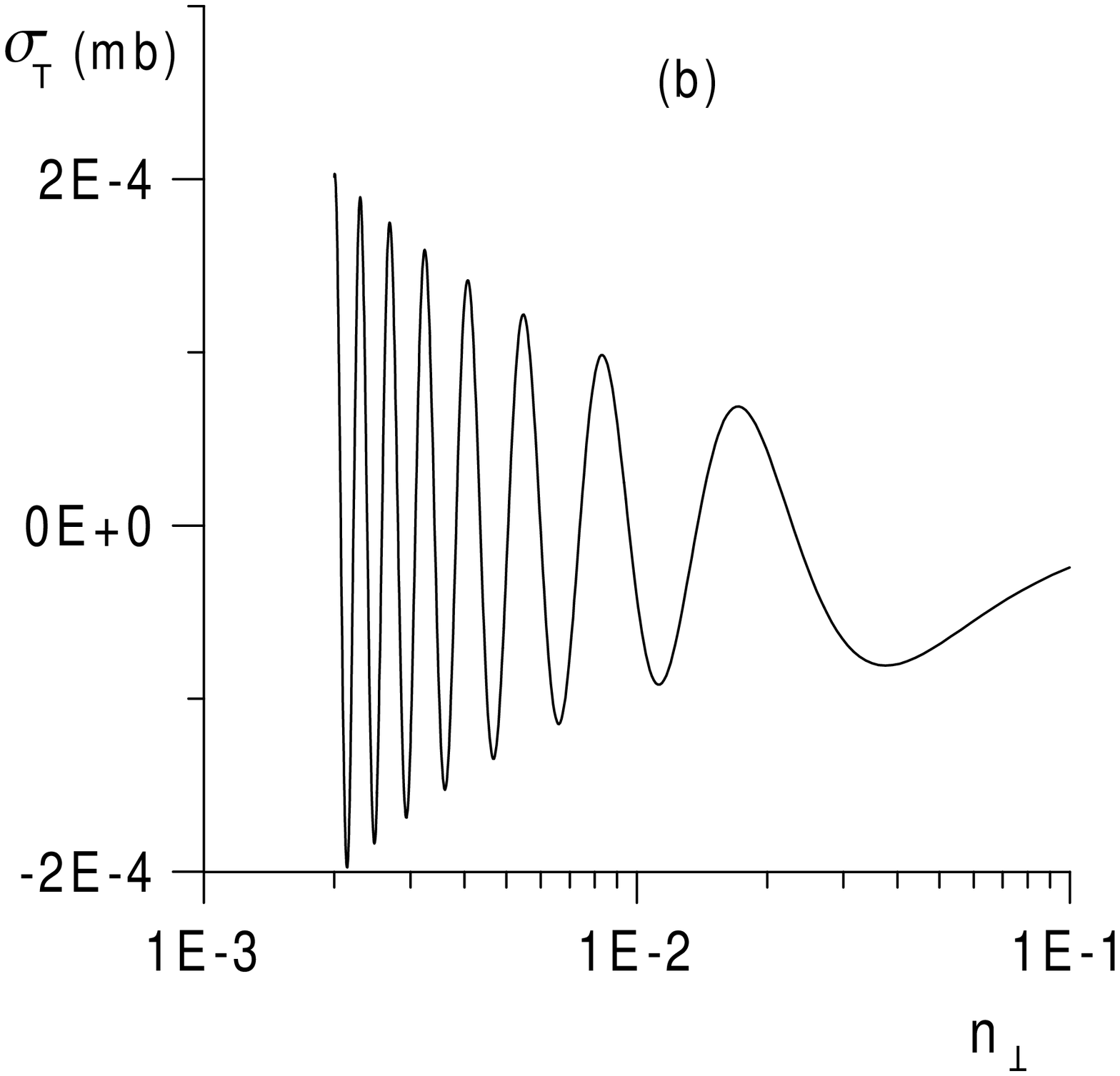}
\caption{
(a)
Angle of T-odd spin rotation of
protons with the energy $ 100 ~\mbox {GeV} $
in tungsten target of
$19.4 ~\mbox {g/cm}^3 $ density,
(b) T-odd cross section (related to one axis atom) of a proton by
(100) tungsten crystal axis .
$n_\bot $ is a perpendicular to the axis component of the unit
Vector in a direction of proton motion.
}
\label {Fig3}
\end {figure}

To simplify the
calculations $g(\rho^2)$ and potential
$v (\rho) $ have been approximated by the simple functions:
\[
g(\rho^2) =g(0) e ^ {-\alpha \rho^2}, ~~
v (\rho) =v (0) e ^ {-\alpha \rho^2},
\]
where  $
{\displaystyle
g(0)=
\frac{Ze}{d}
\frac{3R_a^2}{(R_T^2+R_a^2)R_T^2}
}
$, and $
{ \displaystyle
 v(0)=\frac{Ze^2}{d}\ln\left(\frac{R_T^2+R_a^2}{R_T^2}\right)}$.
For the tungsten the best fitted parameter
$ \alpha $, at $R_T=0.087 ~\mbox {\AA} $ and $R_a=0.27 ~\mbox {\AA} $,
appears to be equal $ \alpha=51.4 ~\mbox {\AA} ^ {-2} $. Distance between
atoms is $d=3.165 ~\mbox {\AA} $ for an axis $ (100) $.
T-odd spin rotation
angle \cite {barbook}
 $ \phi_T $ (per unit particle path length) in tungsten target
and
T-odd cross section by an axis  $ \sigma_T $ (per one axis atom)
for
protons with the energy 100 GeV are shown in fig. \ref{Fig3}.
The
quantities involved
are proportional
to the real and imaginary
parts of $C_1 $ from the expression (\ref {ampl}) accordingly  :
\[
\phi_T =\frac {2\pi \wp {\mbox Re} C_1} {p} n_\bot d, ~~~~
\sigma_T =\frac {2\pi {\mbox Im} C_1} {p} n_\bot d ~.
\]
From the (\ref {end}) it follows, that at large
energies,
when $ \frac {\varepsilon} {p} \approx 1 $,
contribution determined by $ {\mathfrak m} _T $
to the
T-odd amplitude
is proportional to the energy of a particle. So, T-odd
cross section
and angle of spin rotation, appropriate
to this contribution,
 do not depend on energy.
Let's remind, that,
according to the eq.(\ref {vert}),
T-violating charge $ {\mathfrak e} _T $
and magnetic moment $ {\mathfrak m} _T $
are measured in terms of $e $ and $e/2m $ accordingly.
Results shown
in fig.\ref{Fig3} are calculated for
$ {\mathfrak m} _T + {\mathfrak e} _T\left (
1 +\frac {\varepsilon} {m} \right) ^ {-1} =1 $.
For this case the T-odd cross section is about $ 10 ^ {-4} $ mbarn.
Accuracy of cross section measurements
 reaches
$ 10 ^ {-6} $ \cite {COSY} by now, hence, it is possible
to obtain restriction
$ 10 ^ {-2} $ on $ {\mathfrak m} _T $.
Measurements should be made
for angles
$\sim 10^{-2}-10^{-3}$.
In doing so frequency of the T-odd scattering
amplitude oscillation is not high,
in addition,
spin depolarisation (relative decreasing of
absolute
polarisation
magnitude on unit
particle path length)
$\eta=4\wp \frac{Z^2e^4}{m^2}\left(\mu-
\frac{\varepsilon}{\varepsilon+m}\right)^2\frac{R_a-R_T}{d\, n_\bot}$.
is low.
The depolarisation arising from scattering
by atomic chains is calculated
in
the manner described
in
\cite{L} and
comes to $6 ~\%$ in one meter target for the angle $n_\bot=10^{-2}$.

\par
Author is grateful to Dr. K. Batrakov
for discussions,
in which an idea of the work appears,
and to Prof. V. Baryshevsky for valuable
remarks.

\end {document}